\documentclass[11pt]{article}
\usepackage{amsfonts}
\usepackage{amssymb}
\usepackage{amsmath,amssymb}
\usepackage{bbm}

\def\slasha#1{\setbox0=\hbox{$#1$}#1\hskip-\wd0\hbox to\wd0{\hss\sl/\/\hss}}

\def\periodb#1{\setbox0=\hbox{$#1$}#1\hskip-\wd0\hbox to\wd0{-}}

\def\sfrac#1#2{{\textstyle\frac{#1}{#2}}}
\newcommand{\unit}{\mathbbm{1}}   
\newcommand{\frg}{\mathfrak{g}}
\newcommand{\CA}{\mathcal{A}}    
\newcommand{\CG}{\mathcal{G}} 
\newcommand{\CH}{\mathcal{H}} 
\newcommand{\CV}{\mathcal{V}}    
\newcommand{\CL}{\mathcal{L}}    
\newcommand{\CF}{\mathcal{F}}   
\newcommand{\Z}{\mathbb{Z}}    
\newcommand{\R}{\mathbb{R}}     
\newcommand{\C}{\mathbb{C}}     
\newcommand{\Hbb}{\mathbb{H}}     
\newcommand{\Abb}{\mathbb{A}}     
\newcommand{\CPP}{{\mathbb{C}P}}    

\newcommand{\im}{\mathrm{i}} 
 
\newcommand{\dd}{\mathrm{d}}     
\newcommand{\dpar}{\partial}     
\newcommand{\diag}{{\mathrm{diag}}}

\newcommand{\tr}{\mathrm{tr}}     
\newcommand{\+}{\dagger}
\newcommand{\wt}{\widetilde}

\newcommand{\Ad}{\mathrm{Ad}} 
\newcommand{\sU}{\mathrm{U}}     
\newcommand{\sSU}{\mathrm{SU}}

\newcommand{\sGL}{\mathrm{GL}}  
\newcommand{\rsu}{\mathrm{su}}

\newcommand{\veps}{{\varepsilon}} 
\newcommand{\vph}{{{\varphi}}} 
\newcommand{\Ups}{{{\Upsilon}}}

\makeatletter
\renewcommand*\l@section{\@dottedtocline{1}{1.5em}{4em}}
\@addtoreset{equation}{section}
\makeatother

\begin{document}
\begin{titlepage}
\setcounter{page}{0}
.
\vskip 3cm
\begin{center}
{\LARGE \bf Geometric confinement in gauge theories}\\
\vskip 2cm
{\Large Alexander D. Popov}
\vskip 1cm
{\em Institut f\"{u}r Theoretische Physik,
Leibniz Universit\"{a}t Hannover\\
Appelstra{\ss}e 2, 30167 Hannover, Germany}\\
{Email: alexander.popov@itp.uni-hannover.de}
\vskip 1.1cm
\end{center}
\begin{center}
{\bf Abstract}
\end{center}
In 1978, Friedberg and Lee introduced  the phenomenological soliton bag model of hadrons, generalizing the MIT bag model developed in 1974 shortly after the formulation of QCD. In this model, quarks and gluons are confined due to coupling with a real scalar field $\rho$ which tends to zero outside some compact region $S\subset\R^3$ determined dynamically from the equations of motion. The gauge coupling in the soliton bag model is running as the inverse power of $\rho$ already at the semiclassical level. We show that this model arises naturally as a consequence of introducing the warped product metric $\dd s^2_M + \rho^2\dd s^2_G$ on the principal $G$-bundle $P(M,G)\cong M\times G$ with a non-Abelian group $G$ over Minkowski space $M=\R^{3,1}$. Confinement of quarks and gluons in a compact domain $S\subset\R^3$ is a consequence of the collapse of the bundle manifold $M\times G$ to $M$ outside $S$ due to shrinking of the group manifold $G$ to a point. We describe the formation of such regions $S$ as a dynamical process controlled by the order parameter field $\rho$.
\end{titlepage}
\newpage
\setcounter{page}{1}

\section{Introduction}

\noindent
{\bf QCD model.}  Quantum chromodynamics (QCD) was finally formulated by 1973~\cite{Fri, Gross, Polit}.  According to it, the fundamental strong interaction is carried by massless gluons described by the one-form $\CA = \CA_\mu\dd x^\mu$ with values in the Lie algebra $\frg = \rsu(3)$  of the Lie group $G=\sSU(3)$. Here $x^\mu$, $\mu =0,...,3$, are coordinates on Minkowski space $M=\R^{3,1}$ with the metric $\eta =(\eta_{\mu\nu})=\diag (-1, 1, 1, 1)$. The field $\CA$ is a connection on the principal bundle $P(M,G)\cong M\times G$. It is also a connection on the Hermitian vector bundle $E(M,V)=P\times_GV\cong M\times V$ associated with $P$, where $V=\C^3$ is the space of fundamental representation of the group $G=\sSU(3)$. Quarks $\psi$ are spinors on $M$ with values in sections of the complex vector bundle $E$, i.e. they are $\C^3$-valued spinors on $M$. Thus, gluons define geometry of the bundle $E$ affecting quark motion, and quarks are matter fields affecting this geometry. The Lagrangian $\CL_{\sf QCD}$ of QCD is constructed from the curvature $\CF = \dd\CA + \CA\wedge\CA$ of $\CA$, $\psi$ and $\CA$.

\noindent
{\bf MIT bag model.} All hadrons are composed of quarks and neither quarks nor gluons have been observed outside of hadrons (color confinement). In 1974 the bag model was proposed ~\cite{MIT} to describe hadrons. This is a {\it phenomenological} model in which confinement of quarks and gluons is {\it postulated} by imposing boundary conditions such that all fields vanish outside the bag\footnote{In fact, the domain $S$ can have any shape, for example, be an elongated cylinder (flux tube).} $S=\{x\in \R^3\mid r^2=(x^1)^2+(x^2)^2+(x^3)^2\le\R\}.$ In this model the Lagrangian density is chosen in the form
\begin{equation}\label{1.1}
\CL^{}_{\sf bag}=(\CL^{}_{\sf QCD} -\Lambda )\unit_S + \mbox{boundary~terms~on~} \partial S\ ,
\end{equation}
where the constant $\Lambda >0$ is the vacuum energy density inside $S$ and $\unit_S$ is the characteristic function of $S$, 
\begin{equation}\label{1.2}
\unit_S:\ \ \R^3\to \{0,1\}\ ,\  \quad \unit_S(x)=\left\{\begin{array}{l}1\ \mbox{for}\ x\in S\\0\ \mbox{for}\ x\not\in S\end{array}\right .\
\end{equation}
so that $\CL^{}_{\sf bag}(t,x)=0$ for $x\not\in S$. From \eqref{1.1} and \eqref{1.2} it follows that the effective gauge coupling is
\begin{equation}\label{1.3}
\frac{1}{g^2_{\sf eff}}=\frac{1}{g^2_{\sf YM}}\, \unit_S =\left\{\begin{array}{l}\frac{1}{g^2_{\sf YM}}\,\ \mbox{for}\ x\in S \\[3mm]
0\ \mbox{for}\ x\not\in S\end{array}\right . ,
\end{equation}
where $g_{\sf YM}>0$ is the bare coupling constant. It is this profile of $g_{\sf eff}$ is responsible for confinement in the MIT bag model.

\noindent
{\bf Friedberg-Lee bag model.} As a next step, a soliton bag model was proposed~\cite{FrLee}, where $g_{\sf YM}^{-2}\unit_S$ was replaced by a function $g_{\sf YM}^{-2}\rho_S^2$ and $-\Lambda\unit_S$ was replaced by the terms
\begin{equation}\label{1.4}
\CL^{}_{\rho}=-\left (\sfrac12\,\eta^{\mu\nu}\partial_\mu\rho^{}_S\partial_\nu\rho^{}_S + V(\rho^{}_S)\right)\ ,
\end{equation}
where $V(\rho^{}_S)$ is the potential energy for the scalar field $\rho^{}_S$. Also, the coupling of the dynamical scalar field $\rho^{}_S$ with quark fields was introduced. It has been shown that there are solutions with quarks, gluons and $\rho^{}_S$ localized on a compact region $S\subset\R^3$ dynamically determined. This model can be reduced to the MIT bag model~\cite{FrLee}.  The soliton bag model is phenomenological since it is not derived from the first principles and the nature of the field $\rho^{}_S$ is unknown. A description of this and some more sophisticated bag models can be found e.g. in reviews~\cite{Tar, Thom}.

\noindent
{\bf Other models.} The bag models describe the localization of quark and gluons inside hadrons, in a region $S$, but do not describe the mechanism of this localization through a derivation from the fundamental QCD or QCD-like theory. That is why the search of more fundamental models of confinement continued, other models were proposed. A description of the most popular models can be found e.g. in~\cite{Green}. All of these models, over about 50 years of efforts, have failed to produce satisfactory results for QCD. If this is the case, and since the bag models are quite successful phenomenologically,  it makes sense to return to them and derive bag models from a theory more general  than QCD. In fact, our goal is not a confinement per se, but a description of gauge, scalar and fermionic fields with bag type Lagrangians derived from bundle geometry rather than arbitrarily introduced by hands. Confinement will be a consequence.

\section{Conformal extension of Lie group $G$}

\noindent
{\bf Group SU($N$).}  Consider the group\footnote{We often denote SU($N$) as $G$ since all results can be easily generalized to an arbitrary Lie group.} $G=\sSU(N)$ in the defining representation of $N{\times}N$ unitary matrices
\begin{equation}\label{2.1}
\sSU(N)=\left\{g\in \sGL(N, \C)\mid g^\+g=gg^\+=\unit_N  \right\}\ ,
\end{equation}
where
\begin{equation}\label{2.2}
g=\left(g^i_j\right),\quad g^\+=\left(\bar g^{\bar\jmath}_{\bar\imath}\right)\quad\mbox{and}\quad \unit_N=\left(\delta_{i\bar\jmath}\right)=\left(\delta_{\bar\jmath i}\right)
\end{equation}
for $i, j, \bar\imath, \bar\jmath,... = 1,...,N$. In components, the definition \eqref{2.1} reads
\begin{equation}\label{2.3}
g^k_i\,\delta_{k\bar l}\,\bar g^{\bar l}_{\bar\jmath}=\delta_{i\bar\jmath}\ .
\end{equation}
Matrices $g$ act on vectors $\psi\in\C^N$ of fundamental representation of SU$(N)$.

\noindent
{\bf Lie algebra $\rsu(N)$.}  Let $\{I_a\}$ with $a=1,...,N^2-1$ be the generators of SU($N$) in the defining representation, i.e. $I_a$'s are $N{\times}N$ antihermitian matrices $I_a=(I_{aj}^i)$ with the structure constants $f_{ab}^c$ given by the commutation relations
\begin{equation}\label{2.4}
[I_a,I_b]=f^c_{ab}\,I_c\ .
\end{equation}
We can normalize these generators such that $\tr (I_aI_b)=-\delta_{ab}$. They form a basis for the Lie algebra
$\frg =\rsu(N)=\,$Lie$\,\sSU(N)$.

\noindent
{\bf Vector fields and $\rsu(N)$.} Consider the space $\C^N$ with complex coordinates $z^i$, $i=1,...,N$, and their complex conjugate $\bar z^{\bar\imath}$. We introduce on $\C^N$ the holomorphic vector fields
\begin{equation}\label{2.5}
I_a^{\sf v}:= - I_{aj}^i\, z^j\,\partial_i
\end{equation}
with $\dpar_i:=\dpar/\dpar z^i$. They are the image of the generators $I_a$ under the embedding of the algebra $\rsu(N)$ into the algebra of vector fields on $\C^N$ since
\begin{equation}\label{2.6}
[I_a^{\sf v}, I_b^{\sf v}]=f^c_{ab}\,I_c^{\sf v}\ .
\end{equation}
In fact, vector fields  \eqref{2.5} are holomorphic parts of real vector fields $I_a^{\sf v} + (I_a^{\sf v})^\+$ with the same commutators \eqref{2.6}. We use \eqref{2.5} because we act on holomorphic objects.

Consider vector fields on $\C^N$, 
\begin{equation}\label{2.7}
\psi =\psi^i\dpar_i\ ,
\end{equation}
where the components $\psi^i$'s of $\psi=\{\psi^i\}$ do not depend on $z=\{z^i\}\in\C^N$. For commutators of $I_a^{\sf v}$'s with vector fields \eqref{2.7} we have
\begin{equation}\label{2.8}
\delta_a\psi =(\delta_a\psi^i)\dpar_i :=[I_a^{\sf v}, \psi]=  I_{aj}^i\psi^j\dpar_i\ \Rightarrow\   \delta_a\psi^i=I_{aj}^i\psi^j\ .
\end{equation}
It follows from \eqref{2.8} that the vectors \eqref{2.7} belong to the fundamental representation of the group SU$(N)$.

\noindent
{\bf Scalar product.} We introduce on $\C^N$ the Hermitian metric
\begin{equation}\label{2.9}
h=\delta_{i\bar\jmath}\,\dd z^i\dd\bar z^{\bar\jmath}\ ,
\end{equation}
which is invariant under the action of the Lie group SU$(N)$. For two vectors $\psi_1=\psi_1^i\dpar_i$ and $\psi_2=\psi_2^i\dpar_i$ we have
\begin{equation}\label{2.10}
h(\psi_1, \psi_2)=\delta_{i\bar\jmath}\psi_1^i \bar\psi_2^{\bar\jmath}=\psi_2^\+\psi_1
\end{equation}
and hence
\begin{equation}\label{2.11}
\langle\psi , \psi\rangle = h(\psi , \psi )=\delta_{i\bar\jmath}\psi^i\bar\psi^{\bar\jmath}=\psi^\+\psi
\end{equation}
for components $\psi=\{\psi^i\}$ in the basis $\{\dpar_i\}$. 
The group SU$(N)$ acts on $\psi\in\C^N$ by the rule
\begin{equation}\label{2.12}
\psi\mapsto\tilde\psi = g\psi , \quad \tilde\psi^i=g^i_j\psi^j\quad\Rightarrow\quad\tilde\psi^\+=\psi^\+g^\+
\end{equation}
and \eqref{2.11} is invariant under these transformations.

\noindent
{\bf Group manifold.} Consider group elements $g\in G=\sSU(N)$ and introduce left- and right-invariant one-forms on the Lie group $G$ considered as a smooth manifold,
\begin{equation}\label{2.13}
g^{-1}\dd g=:\theta^a_LI_a\quad\mbox{and}\quad (\dd g)g^{-1}=:\theta^a_RI_a\ ,
\end{equation}
where $\dd$ is the exterior derivative on $G$ and $I_a=(I^i_{aj})$. For metric on $G$ we have
\begin{equation}\label{2.14}
\dd s^2_G=\delta_{ab}\theta^a_L\theta^b_L=\delta_{ab}\theta^a_R\theta^b_R\ ,
\end{equation}
where 
\begin{equation}\label{2.15}
\theta^a_R=D^a_b\theta^b_L\quad\mbox{for}\quad gI_ag^{-1}=:D_a^bI_b\ .
\end{equation}
From \eqref{2.15} one can see that left- and right-invariant objects are interchangeable. 

The forms $\theta^a_L$ obey the Maurer-Cartan equations
\begin{equation}\label{2.16}
\dd\theta_L^a + \sfrac12\,f^a_{bc}\theta_L^b\wedge\theta_L^c=0
\end{equation}
and the same equations for $\theta^a_R$ with $f^a_{bc}\to -f^a_{bc}$. We introduce left- and right-invariant vector fields on $G$ dual to $\theta^a_L$ and $\theta^a_R$,
\begin{equation}\label{2.17}
L_a\lrcorner\,\theta^b_L = \delta_{a}^b\quad\mbox{and}\quad R_a\lrcorner\,\theta^b_R = \delta_{a}^b\ ,
\end{equation}
which obey the equations
\begin{equation}\label{2.18}
[L_a, L_b] = f^c_{ab}L_c\quad\mbox{and}\quad [R_a, R_b] = - f^c_{ab}R_c
\end{equation}
and commute with each other.

\noindent
{\bf Adjoint representation.}  The center of $\sSU(N)$ is given by the matrices $\zeta\unit_N$, where $\zeta$ is the $N$-th root of unity, $\zeta^N=1$, i.e.
\begin{equation}\label{2.19}
Z(\sSU(N)) = \Z/N\Z =:\Z_N\ .
\end{equation}
Consider the left action of $G$ on itself,
\begin{equation}\label{2.20}
G\ni f\ \ \mapsto\ \ f^g:= gfg^{-1}\in G \quad\mbox{for}\quad f,g\in G\ .
\end{equation}
The maps \eqref{2.20} are inner automorphisms of $G$ denoted Inn$(G)$, and we have an isomorphism
\begin{equation}\label{2.21}
G/Z(G)\cong \mbox{Inn}(G)\ \Rightarrow\ \mbox{Inn}(\sSU(N))=\sSU(N)/\Z_N\ .
\end{equation}
The group PU$(N)=\sSU(N)/\Z_N$ (projective unitary group) has no $N$-dimensional representations.  The adjoint action \eqref{2.20} of PSU$(N)$ on $\sSU(N)$ induces the action
\begin{equation}\label{2.22}
\Ad_g:\ \ T_e G\to T_e G, \quad \frg\ni\vph\ \mapsto\ g\vph g^{-1}\in \frg\ ,
\end{equation}
where $\frg = T_e G =\rsu(N)$ is the tangent space to $G$ at the origin $e\in G$. Thus, the group $\mbox{PU}(N)$ has $(N^2-1)$-dimensional representation \eqref{2.22} which is the adjoint representation of $\sSU(N)$. Matrices $D=(D^a_b)$ introduced in \eqref{2.15} are matrices of this representation $g\to D$. For  fields $\vph =\vph^aI_a\in\frg$ in the adjoint representation we have
\begin{equation}\label{2.23}
\widetilde\vph = \widetilde\vph^aI_a:=g\vph g^{-1}=(gI_b g^{-1})\vph^b=D^a_b\vph^bI_a\ \ \Rightarrow\ \ \widetilde\vph^a=D^a_b\vph^b\ ,
\end{equation}
i.e. they are transformed with the matrices $D$. The metric \eqref{2.14} is invariant under these $D$-rotations.

\noindent
{\bf Lie group $\widetilde G$.} Consider the conformal extension of the group $G$ defined as~\cite{Kob}
\begin{equation}\label{2.24}
\widetilde G=\R^+\times G=\bigl\{\tilde g = \rho g, \ \rho\in\R^+, g\in G \mid \tilde g^\+\tilde g =\rho^2\unit_N\bigr\}\ ,
\end{equation}
where $\R^+=\sGL^+(1,\R)$ is the multiplicative group of positive real numbers.
As a manifold, the group \eqref{2.24} is a cone $C(G)$ over $G$ with the metric
\begin{equation}\label{2.25}
\dd s^2_{\widetilde G}=\sigma_0^2(\dd\rho^2 + \rho^2\dd s^2_G)\ ,
\end{equation}
where we introduced a constant factor $\sigma_0^2$ for future convenience.
If we add the tip $\tilde g =0$ to the cone $C(G)$ then we get a semigroup with identity (a monoid), since the element $\tilde g=0$ has no inverse.

The metric \eqref{2.25} can be written as
\begin{equation}\label{2.26}
\dd s^2_{\widetilde G}=\sigma^2_0\, e^{2\tau} (\dd\tau^2 + \dd s^2_G) = g_{\tau\tau}\dd\tau^2 + g_{ab}\theta^a_L\theta^b_L\ ,
\end{equation}
where 
\begin{equation}\label{2.27}
g_{\tau\tau}^{}=\sigma_0^2\, e^{2\tau}\quad\mbox{and}\quad
g_{ab}^{}=\sigma_0^2\, e^{2\tau}\delta_{ab}
\quad\mbox{for}\quad\tau=\ln\rho\ ,
\end{equation}
and
\begin{equation}\label{2.28}
\dd s^2_{\R\times G}=\dd\tau^2 + \dd s^2_G\quad\mbox{for}\quad\tau\in \R
\end{equation}
is the metric on the cylinder. Introducing the vector field $\dpar_\tau$ dual to the one-form $\dd\tau$, we obtain a basis $\{\dpar_\tau , L_a\}$ or $\{\dpar_\tau , R_a\}$ of the Lie algebra $\tilde\frg =\,$Lie$\,\widetilde G$ realized as vector fields on $\widetilde G$.

\noindent
{\bf Algebra $\tilde\frg$ on $\C^N$.} We described the Lie algebra $\tilde\frg =\R\oplus\frg$ via vector fields on the group $\widetilde G$ considered as a manifold. On the other hand, we mapped $\frg = \rsu(N)$ to the holomorphic parts $I_a^{\sf v}$ of real vector fields $I_a^{\sf v}+(I_a^{\sf v})^\+$ on $\C^N$. Extension to $\tilde \frg$ is given by the generator
\begin{equation}\label{2.29}
I_\tau^{\sf v}=-I^i_{\tau j}z^j\dpar_i=-z^i\dpar_i\quad\mbox{with}\quad I^i_{\tau j}=\delta^i_j
\end{equation}
commuting with $I_a^{\sf v}$. It is easy to see that
\begin{equation}\label{2.30}
\delta_\tau\psi =[I_{\tau}^{\sf v}, \psi]=\psi^i\dpar_i=\psi\ ,
\end{equation}
i.e. $\exp(\alpha\delta_\tau)\psi = \exp(\alpha)\psi$ (dilatations). The Hermitian metric \eqref{2.11} is not invariant under these transformations,
\begin{equation}\label{2.31}
h(\psi , \psi)=\psi^\+\psi\quad\mapsto\quad\tilde h=\tilde\psi^\+\tilde\psi =e^{2\alpha}h\ ,
\end{equation}
i.e. \eqref{2.30} generates a conformal transformation of the Hermitian metric \eqref{2.9} on $\C^N$. This metric is invariant under rotations but not under conformal transformations.

\section{Gauge theory with the structure group $\widetilde G$}

\noindent
{\bf Principal bundle  $\wt P$.} We consider Minkowski space $M=\R^{3,1}$ with the metric
\begin{equation}\label{3.1}
\dd s^2_M=\eta^{}_{\mu\nu}\dd x^\mu\dd x^\nu,\quad \eta =(\eta^{}_{\mu\nu})=\diag (-1, 1, 1, 1)\ ,
\end{equation}
where $x^\mu$ are coordinates on $M$, $\mu ,\nu = 0,...,3$. Consider the direct product
\begin{equation}\label{3.2}
\wt P=M\times \wt G
\end{equation}
of $M$ and the group $\wt G=C(G)$ introduced in Section 2 for $G=\sSU(N)$. On the manifold \eqref{3.2} one can introduce a direct product metric
\begin{equation}\label{3.3}
\dd s^2_{M\times\wt G}=\dd s^2_M + \dd s^2_{\wt G}\ ,
\end{equation}
where $\dd s^2_{\wt G}$ is written down in \eqref{2.25}-\eqref{2.27}.

The metric \eqref{3.3} can be twisted so that the manifold $\wt P$ with the twisted metric will be a direct product \eqref{3.2} only topologically but not as a smooth manifold. This is done with the help of a connection one-form $\tilde\CA =\tilde\CA_\mu\dd x^\mu$ on $M$ with values in the Lie algebra $\wt\frg =\,$Lie$\,\wt G=\R\oplus \rsu(N)$.

\noindent
{\bf Frame and coframe on $\wt P$.} Consider the one-form $\tilde\CA =\tilde\CA_\mu\dd x^\mu$ on $M$ whose components are vector fields on $\wt G$,
\begin{equation}\label{3.4}
\tilde\CA_\mu = \CA_\mu^aV_a +  \CA_\mu^\tau V_\tau = \CA_\mu^a (-R_a) + \CA_\mu^\tau \dpar_\tau\ ,
\end{equation}
where $\CA_\mu^a$ and $a_\mu := \CA_\mu^\tau $ depend only on $x\in M$. Introduce vector fields
\begin{equation}\label{3.5}
\wt\nabla_\mu =\nabla_\mu + a_\mu V_\tau = \dpar_\mu + \CA_\mu^aV_a + a_\mu V_\tau ,\quad
\wt\nabla_\tau =V_\tau\quad\mbox{and}\quad \wt\nabla_a=L_a\ ,
\end{equation}
and dual one-forms
\begin{equation}\label{3.6}
\Theta^\mu =\dd x^\mu ,\quad \Theta^\tau =\dd\tau - a_\mu \dd x^\mu\quad\mbox{and}\quad
\Theta^a=\theta^a_L + D^{-1a}_{~~~b}\,\CA^b_\mu\,\dd x^\mu \ ,
\end{equation}
where $L_a, R_a, \theta^a_L, \theta^a_R$ and $D^a_b$ were introduced in Section 2. It is easy to check that
\begin{equation}\label{3.7}
\begin{array}{lll}
\wt\nabla_\mu \lrcorner\Theta^\nu =\delta^\nu_\mu ,&  \wt\nabla_\mu \lrcorner\Theta^\tau =0,  &  \wt\nabla_\mu \lrcorner\Theta^a =0, \\
\wt\nabla_\tau  \lrcorner\Theta^\nu =   0, &   \wt\nabla_\tau  \lrcorner\Theta^\tau =1, &\wt\nabla_\tau  \lrcorner\Theta^a =   0, \\
\wt\nabla_a  \lrcorner\Theta^\nu =    0, &\wt\nabla_a \lrcorner\Theta^\tau =   0, &\wt\nabla_a \lrcorner\Theta^b =\delta_a^b, 
\end{array}
\end{equation}
where ``$\lrcorner$" is the internal product of vector fields and one-forms. Vector fields \eqref{3.5} form a frame on the tangent bundle of $\wt P$ and one-forms \eqref{3.6} form a coframe.

\noindent
{\bf Metric on $\wt P$.} Using the coframe \eqref{3.6}, we can introduce the metric
\begin{equation}\label{3.8}
\dd s^2_{\widetilde P}= \dd s^2_{M} +  (g_{\tau\tau}\Theta^\tau\Theta^\tau +g_{ab}\Theta^a\Theta^b)\ ,
\end{equation}
where $g_{\tau\tau}$ and $g_{ab}$ are given in \eqref{2.27}. This metric reduces to the direct product metric \eqref{3.3} for $\tilde\CA =0$. The manifold \eqref{3.2} with the metric \eqref{3.8} is a trivial principal bundle $\wt P(M, \wt G)$ with the structure group $\wt G$ and a connection $\tilde\CA$. Note that if we put $\tau =0\ \Leftrightarrow\ \rho =1$ and $a_\mu =0$ then the group $\wt G$ reduces to $G$, $\Theta^\tau =0$ and the bundle $\wt P$ reduces to the principal bundle $P(M, G)=M\times G$ with a connection $\CA =\CA^a_\mu\dd x^\mu V_a$.

\noindent
{\bf Curvature $\wt\CF$ of $\tilde\CA$.} Consider the covariant derivatives \eqref{3.5}. Calculating commutators of these vector fields, we obtain
\begin{equation}\label{3.9}
\wt\CF_{\mu\nu}:=[\wt\nabla_\mu , \wt\nabla_\nu ]=[\dpar_\mu + \tilde\CA_\mu , \dpar_\nu + \tilde\CA_\nu]=\CF^a_{\mu\nu}V_a + f_{\mu\nu}V_\tau\ ,
\end{equation}
\begin{equation}\label{3.10}
\wt\CF_{\mu\tau}=[\wt\nabla_\mu , \wt\nabla_\tau ]=0,\ \
\wt\CF_{\mu a}=[\wt\nabla_\mu , \wt\nabla_a ]=0,\ \
\wt\CF_{\tau a}=[\wt\nabla_\tau , \wt\nabla_a]=0,
\end{equation}
\begin{equation}\label{3.11}
\wt\CF_{ab}:=[\wt\nabla_a, \wt\nabla_b] - f^c_{ab}\wt\nabla_c=0,
\end{equation}
where 
\begin{equation}\label{3.12}
\CF_{\mu\nu}^a=\dpar_\mu\CA_\nu^a - \dpar_\nu\CA_\mu^a + f^a_{bc}\CA_\mu^b\CA_\nu^c\quad\mbox{and}\quad f_{\mu\nu}=\dpar_\mu a_\nu - \dpar_\nu a_\mu\ .
\end{equation}
The two-form
\begin{equation}\label{3.13}
\wt\CF = \sfrac12\, \CF_{\mu\nu}^a\,\dd x^\mu\wedge\dd x^\nu V_a+ \sfrac12\, f_{\mu\nu}\,\dd x^\mu\wedge\dd x^\nu V_\tau
\end{equation}
with values in the Lie algebra $\wt\frg = \frg\oplus\R$ is the curvature of the connection $\tilde\CA$ on $\wt P$. The standard matrix form of connection and curvature,
\begin{equation}\label{3.14}
\tilde\CA_\mu = \CA_\mu^a I_a + a_\mu I_\tau\quad\mbox{and}\quad\wt\CF_{\mu\nu}=\CF_{\mu\nu}^a I_a +  f_{\mu\nu}I_\tau\ ,
\end{equation}
appears after substitution
\begin{equation}\label{3.15}
V_a\to I_a=(I_{aj}^{~i})\quad\mbox{and}\quad V_\tau\to I_\tau =\unit_N=(\delta^i_j)\ .
\end{equation}
This appears due to dependence on $\wt G$-variables via $\tilde g^{-1}=e^\tau g^{-1}$ and identities $V_a\tilde g^{-1}=\tilde g^{-1}I_a$, $V_\tau\tilde g^{-1}=\tilde g^{-1}$.

\noindent
{\bf Vector bundle $\wt E$.} The fibres of the principal bundle $\wt P\to M$ over points $x\in M$ are groups $\wt G_x\cong C(\sSU(N))$ of $N{\times}N$ matrices defined in \eqref{2.24}. Now their elements depend on $x\in M$ and we denote them by $\Phi^{-1}=\rho\phi^{-1}=\rho\phi^\+\in \R^+\times G$ with functions $\rho\in\R^+$ and $\phi\in G$. In the definition \eqref{2.24} with $x$-dependence we use $\tilde g=\Phi^{-1}$ so that $\wt G$ acts on the vector space $V=\C^N$ as $\xi\mapsto\Phi^{-1}\xi$ for $\xi\in V$ (right action). We associate with $\wt P$ the complex vector bundle
\begin{equation}\label{3.16}
\wt E=\wt P\times_{\wt G}V=\{\wt P\times V\ni (\tilde p, \xi)\sim (\tilde p\Phi , \Phi^{-1}\xi )\in\wt P\times V\}\cong M\times V
\end{equation}
endowed with the Hermitian metric  \eqref{2.9}-\eqref{2.11} on fibres $\wt E_x=V_x\cong\C^N$. On fibres $\wt E_x$ of $\wt E$ we have vector fields \eqref{2.5} and \eqref{2.29}. Sections of $\wt E$ are given by vector fields \eqref{2.7} with $\psi^i =\psi^i(x)$ depending on $x\in M$.

\noindent
{\bf Frames and coframes on $\wt E$.} On the manifold \eqref{3.16} we introduce vector fields
\begin{equation}\label{3.17}
\hat\nabla_\mu = \dpar_\mu + \CA_\mu^a I_a^{\sf v}+a_\mu I_\tau^{\sf v}\ ,\quad \hat\nabla_i=I^{\sf v}_i:=\dpar_i\ ,
\end{equation}
and one-forms
\begin{equation}\label{3.18}
\Theta^\mu =\dd x^\mu\ ,\quad \Theta^i=\dd z^i+\CA_\mu^a\dd x^\mu I_{aj}^{~i}z^j +a_\mu\dd x^\mu I_{\tau j}^{~i}z^j\ ,
\end{equation}
where $\CA_\mu^a$ and $a_\mu$ are components of the connection $\tilde\CA$ on the principal bundle $\wt P$. It is easy to see that
\begin{equation}\label{3.19}
\hat\nabla_\mu \lrcorner\Theta^\nu =\delta_\mu^\nu ,\ \hat\nabla_\mu \lrcorner\Theta^i =0,\ \hat\nabla_i \lrcorner\Theta^\mu =0\quad\mbox{and}\quad\hat\nabla_i \lrcorner\Theta^j=\delta^j_i ,
\end{equation}
i.e. vector fields \eqref{3.17} are dual to one-forms \eqref{3.18}.

\noindent
{\bf Metric on $\wt E$.} Vector fields \eqref{3.17} form a frame on tangent bundle of $\wt E$ and one-forms \eqref{3.18} form a coframe. On $\wt E$ we can introduce the metric
\begin{equation}\label{3.20}
\dd s^2_{\wt E}=\dd s^2_M+\rho^2\delta_{i\bar\jmath}\Theta^i\bar\Theta^{\bar\jmath}\ .
\end{equation}
The metric on fibres ${\wt E}_x\cong\C^N$ of $\wt E$ for any fixed $x\in M$ is conformal to the metric \eqref{2.9},
\begin{equation}\label{3.21}
\dd s^2_{\wt E|{\wt E}_x}=\tilde h=\rho^2h=\rho^2\delta_{i\bar\jmath}\dd z^i{\dd \bar z}^{\bar\jmath}\quad\mbox{for}\quad \tilde h_{i\bar\jmath}=\rho^2\delta_{i\bar\jmath}\ ,
\end{equation}
where $\rho = e^\tau$.

\noindent
{\bf Curvature.}  Calculating commutators of the vector fields \eqref{3.17}, we obtain
\begin{equation}\label{3.22}
\hat\CF_{\mu\nu}=[\hat\nabla_\mu , \hat\nabla_\nu ]=\CF_{\mu\nu}^aI_a^{\sf v}+f_{\mu\nu}I_\tau^{\sf v}\ ,\quad \hat\CF_{ij}=[\hat\nabla_i , \hat\nabla_j ]=0\ ,
\end{equation}
\begin{equation}\label{3.23}
\hat\CF_{\mu i}=[\hat\nabla_\mu , \hat\nabla_i ]=\tilde\CA_{\mu i}^{~j}I_j^{\sf v}=
(\CA_\mu^aI_{a i}^{~j}+a_\mu I_{\tau i}^{~j})I_j^{\sf v}\ ,
\end{equation}
where $\CF_{\mu\nu}^a$ and $f_{\mu\nu}$ are written down in \eqref{3.12}, and vector fields $I_a^{\sf v}$ and $I_\tau^{\sf v}$ are defined in \eqref{2.5} and \eqref{2.29}. The vector field $I_i^{\sf v}=\dpar_i$ are generators of translations on $V=\C^N$. Together with $I_a^{\sf v}$ and $I_\tau^{\sf v}$ they form generators of inhomogeneous group $I\wt G$ which is the semidirect product
\begin{equation}\label{3.24}
I\wt G = V\rtimes\wt G
\end{equation}
of groups $V$ and $\wt G$.

The two-form on $M$,
\begin{equation}\label{3.25}
\hat\CF=\sfrac12\,\CF_{\mu\nu}^a\,\dd x^\mu\wedge\dd x^\nu\,I_a^{\sf v}+\sfrac12\,f_{\mu\nu}\,\dd x^\mu\wedge\dd x^\nu\,I_\tau^{\sf v}
\end{equation}
is the curvature of the connection
\begin{equation}\label{3.26}
\hat\CA = \CA^a_\mu\dd x^\mu\,I_a^{\sf v}+a_\mu\dd x^\mu\,I_\tau^{\sf v}
\end{equation}
on the bundle $\wt E$. Both connection and curvature take values in the Lie algebra $\wt\frg = \rsu (N)\oplus\R$. 

Let us consider a section $\psi=\psi^i\dpar_i=\psi^iI_i^{\sf v}$ of the bundle $\wt E$ with components $\psi^i=\psi^i(x)$ depending on $x\in M$. It follows from \eqref{3.23} that
\begin{equation}\label{3.27}
[\hat\nabla_\mu , \psi]=[\dpar_\mu + \hat\CA_\mu , \psi] = (\dpar_\mu\psi^i + \CA^a_\mu I_{aj}^i\psi^j + a_\mu\delta^i_j\psi^j)\dpar_i\ ,
\end{equation}
i.e. the covariant derivative of $\C^N$-valued functions $\psi$ on $M$ is, in fact, components $\hat\CF_{\mu\psi}=[\hat\nabla_\mu ,\hat\nabla_\psi ]$ of the curvature \eqref{3.23}. Also, from \eqref{3.27} it follows that in \eqref{3.25}-\eqref{3.26} we can substitute $I_a^{\sf v}\to I_a$, $I^{\sf v}_\tau\to\unit_N$ and obtain the matrix form \eqref{3.14} for $\hat\CA$ and $\hat\CF$.

\noindent
{\bf Automorphisms of $\wt E$.} Recall that the bundle $\wt P$ is the bundle of conformal frames on the complex vector bundle $\wt E$. All such frames are parametrized by the group
\begin{equation}\label{3.28}
\wt\CG :=\CG_{\wt G}=C^\infty (M, \wt G)
\end{equation}
which has two subgroups - the subgroup
\begin{equation}\label{3.29}
\CG_{G}=C^\infty (M, G)\subset \wt\CG
\end{equation}
of $G$-rotations on fibres $\wt E_x$ of $\wt E$ and the subgroup
\begin{equation}\label{3.30}
\CG_{\R^+}=C^\infty (M, \R^+)\subset \wt\CG
\end{equation}
of conformal transformations of metric on fibres of $\wt E$.

For a matrix $\Ups = u^{-1}g\in \R^+\times G$ the frame $\{\dpar_i\}$ is transformed as
\begin{equation}\label{3.31}
\dpar_i\mapsto\dpar_i^\Ups =\Ups_i^j\dpar_j\quad\Rightarrow\quad\dd z^i\mapsto (\dd z^i)_\Ups=
\Ups^{-1i}_{~~j}\dd z^j\ .
\end{equation}
For the coframe $\Theta^i$ in \eqref{3.18} we have
\begin{equation}\label{3.32}
\Theta^i\mapsto\Theta^i_\Ups = (\dd z^i)_\Ups + \tilde\CA^{\Ups i}_{~~j} (\dd z^j)_\Ups =
u((\dd z)^i_g + \CA^{gi}_{~j}z^j_g + a^u z^i_g),
\end{equation}
where 
\begin{equation}\label{3.33}
(\dd z^i)_g = g^{-1i}_{~~j}\dd z^j ,\quad z^i_g =g^{-1i}_{~j} z^j 
\end{equation}
and
\begin{equation}\label{3.34}
\tilde\CA^\Ups = \Ups^{-1}\tilde\CA\Ups + 
\Ups^{-1}\dd\Ups = (g^{-1}\CA g + g^{-1}\dd g)+(a-\ln u)\unit_N=: \CA^g + a^u\unit_N
\end{equation}
is the $\Ups$-transformed connection $\tilde \CA =\CA + a\unit_N$ on $\wt E$. Formula \eqref{3.34} can be obtained from \eqref{3.27} for $\psi =\psi^i\dpar_i=\psi^i_\Ups\dpar^\Ups_i$ with
\begin{equation}\label{3.35}
\psi^i_\Ups =\Ups^{-1i}_{~~~j}\psi^j
\end{equation}
and $\dpar^\Ups_i$ given in \eqref{3.31}. It follows from \eqref{3.34} that the group $\wt\CG$ acts on $\tilde\CA$ as follows:
\begin{equation}\label{3.36}
\CG_G:\quad \CA\ \mapsto\ \CA^g=g^{-1}\CA g+g^{-1}\dd g,\quad a\mapsto a\ ,
\end{equation}
\begin{equation}\label{3.37}
\CG_{\R^+}:\quad \CA\ \mapsto\ \CA , \quad a\mapsto a^u=a-\dd\ln u\ ,
\end{equation}
i.e. $\CG_G$ acts only on $\CA$ and $\CG_{\R^+}$ acts only on $a$. For $\wt\CF$ from \eqref{3.14} we have
\begin{equation}\label{3.38}
\CF_{\mu\nu}\mapsto\CF_{\mu\nu}^g=g^{-1}\CF_{\mu\nu}g\quad\mbox{and}\quad f_{\mu\nu}\mapsto f_{\mu\nu}^u=f_{\mu\nu}
\end{equation}
These formulae \eqref{3.36}-\eqref{3.38} define the usual transformations under the action of automorphism group $\wt\CG=\,$Aut$_{\wt G}\wt E$. It follows from \eqref{3.31}-\eqref{3.33} that this group does not preserve the metric \eqref{3.21} on fibres $\wt E_x$ of the bundle $\wt E$. For $\Ups=u^{-1}g\in\R^+\times G=\wt G$ we have $h\mapsto u^2h$ and $g$-rotations preserve $h$.

\noindent
{\bf Equations for $\tilde A$.} We considered smooth complex vector bundle $\wt E(M, V)$ over $M$ and the principal bundle $\wt P(M,\wt G)$ of conformal frames on $\wt E$. We have a connection $\tilde\CA$ on both of these bundles. It should be emphasized that the connection $\tilde\CA$ does not have to satisfy any differential equations. In fact, connection \eqref{3.4} on $\wt P$ defines a splitting of tangent spaces $T_p\wt P$ at $p\in\wt P$ into horizontal $\CH_p$ and vertical $\CV_p$ subspaces, $T_p\wt P=\CH_p\oplus\CV_p$. To see this, we should simply rewrite the exterior derivative $\dd_{\wt P}$ on $\wt P$ as follows:
\begin{equation}\label{3.39}
\begin{split}
\dd_{\wt P}&=\dd^{}_{M}+\dd^{}_{\wt G}=\dd x^\mu\dpar_\mu + \dd\tau\dpar_\tau + \theta^a_L L_a=
(\dd^{}_{M}  + \tilde\CA)+(\dd^{}_{\wt G}-\tilde\CA )     \\ 
&=\dd x^\mu (\dpar_\mu + \tilde \CA_\mu ) + (\dd\tau - a)\dpar_\tau + (\theta^a_L + 
D^{-1a}_{~~~b}\CA^b)L_a=\Theta^\mu\wt\nabla_\mu + \Theta^\tau\dpar_\tau + \Theta^aL_a\ .
\end{split}
\end{equation}
This splitting gives us the frame \eqref{3.5} and coframe \eqref{3.6}. Similarly, for the exterior derivative $\dd_{\wt E}$ on $\wt E$ we have
\begin{equation}\label{3.40}
\dd_{\wt E}=\dd_M+\dd_V=\dd x^\mu\dpar_\mu + \dd z^i\dpar_i=\Theta^\mu\hat\nabla_\mu + \Theta^i\hat\nabla_i\ ,
\end{equation}
and this defines vertical subspaces $\wt E_x$ in $\wt E$. 

We repeat once again that, generally speaking, there are no restrictions on the connection $\tilde\CA$ except for the smoothness class. Restrictions on bundles can induce restrictions on connections. For instance, if the bundle is locally constant then $\tilde\CA$ must be flat; if $\wt E$ is holomorphic then the (0,2)-components of the curvature of $\tilde\CA$ have to vanish. In contrast to this, the Yang-Mills (YM) equations on Minkowski space impose restrictions on connections that has nothing to do with geometry. The YM equations are arbitrariness that can only be justified by experimental data. Possible modifications of these equations, derivable from the geometry of the bundle $\wt E$, will be considered in the next section.

\section{Lagrangians for gauge fields, scalars and fermions}

\noindent
{\bf Standard Yang-Mills.} In Section 3 we discussed how the geometry of vector bundles $\wt E$ is characterized by connections $\tilde \CA =(\CA , a)$ and conformal frame $\Phi =(\phi ,\rho )$ parametrized by the Lie group $\wt\CG = C^\infty (M, \wt G)$. We denote by $\tilde\Abb$ the space of all smooth connections $\tilde\CA$ on $\wt E$ and by $\tilde\Abb^\Phi :=\tilde\Abb\times\wt\CG$ the space of pairs $(\tilde\CA , \Phi )$, $\tilde\CA\in \tilde\Abb$ and $\Phi\in\wt\CG$. The standard Lagrangian for massless gauge fields is defined on the tangent bundle $T\tilde\Abb$ and we start from the case $\rho =1$ and $\phi =\unit_N$, which is the ``standard" case. Then the Lagrangian density for the curvature $\wt\CF = \CF + f\unit_N$ is
\begin{equation}\label{4.1}
\CL_{\sf YM}=-\frac{1}{4g^2_*}\,\tr\wt\CF^\+_{\mu\nu}\wt\CF^{\mu\nu}=-\frac{1}{4g^2_*}\,\left(\tr\CF^\+_{\mu\nu}\CF^{\mu\nu} + Nf_{\mu\nu}f^{\mu\nu}\right )\ ,
\end{equation}
where we identify $\sigma_0^{-1}$ from \eqref{2.26}, \eqref{2.27} and \eqref{3.8} with the gauge coupling constant $g_*\equiv g_{\sf YM}>0.$ The factor $N$ in \eqref{4.1} is related to the embedding $a_\mu$ in $N{\times}N$ matrices as $a_\mu\unit_N$.

Let us now act on $\tilde\CA\in\tilde\Abb$ in \eqref{4.1} by the element $\Phi =\rho^{-1}\phi\in\wt\CG$ according to the formula \eqref{3.34} with $\Ups =\Phi$, taking into account that $\CG_{\R^+}$ acts on the metric on fibres $\wt E_x$ of $\wt E$. It is easy to see that \eqref{4.1} is transformed to 
\begin{equation}\label{4.2}
\CL^\rho_{\sf YM}= - \frac{\rho^2}{4g^2_*}\,\left(\tr(\CF^\phi_{\mu\nu})^\+\CF^{\mu\nu}_\phi + Nf_{\mu\nu}^\rho f^{\mu\nu}_\rho\right )= - \frac{\rho^2}{4g^2_*}\,\left(\tr(\CF^\+_{\mu\nu}\CF^{\mu\nu}) + Nf_{\mu\nu}f^{\mu\nu}\right )\ .
\end{equation}
The Lagrangian density \eqref{4.2} depends on $\rho$ but does not depend on $\phi\in\CG_G$, i.e. it does not depend on the choice of SU$(N)$-frame on $\wt E$. However, $\rho$ in \eqref{4.2} is nondynamical and the only reasonable choice is $\rho =1$ and $a_\mu =0=f_{\mu\nu}$. Then \eqref{4.2} reduces to the standard Yang-Mills Lagrangian for the Lie group $G=\sSU (N)$. By virtue of the invariance of \eqref{4.2} under the action of the group $\CG_G=C^\infty (M,G)$, the dynamics of the theory \eqref{4.2} with $\rho =1$ descends to the manifold $\Abb/\CG_G$, where $\Abb$ is the space of smooth connections $\CA$ on the bundle $E=\wt E^{}_{|\rho =1}$.

\noindent
{\bf Scalar fields.} In Section 3, we saw that the geometry of the bundle $\wt E$ is described not only by the curvature \eqref{3.22}, but also by the connection $\tilde\CA$ included in the metric \eqref{3.20} and curvature \eqref{3.23}. Therefore, it is reasonable to add the mass term
\begin{equation}\label{4.3}
\CL_m=-\sfrac12\,v^2_*\,\eta^{\mu\nu}\,\tr(\tilde\CA^\+_\mu\tilde\CA_\nu )=-\sfrac12\,v^2_*\,\eta^{\mu\nu}\left (\tr(\CA^\+_\mu\CA_\nu )+N a_\mu a_\nu\right )
\end{equation}
with an arbitrary parameter $v_*\ge 0$ and a mass $m=v_*g_*$.

Let us now act on $\tilde\CA\in\tilde\Abb$ in \eqref{4.3} by the element $\Phi =\rho^{-1}\phi\in\wt\CG =\,$Aut$_{\wt G}\wt E$ according to the formula \eqref{3.34}. For $\Ups =\Phi$ in \eqref{3.34} we obtain
\begin{equation}\label{4.4}
\begin{split}
\CL_m\mapsto\CL_m^\Phi &=-\sfrac12\,v^2_*\rho^2\,\eta^{\mu\nu}\,\tr(\tilde\CA^\Phi_\mu)^\+\CA_\nu^\Phi 
\\
&=-\sfrac12\,v^2_*\rho^2\,\tr(\nabla_\mu\phi)^\+\nabla^\mu\phi - 
\sfrac{N}{2}\,v^2_*(\dpar_\mu\rho -a_\mu\rho) (\dpar^\mu\rho -a^\mu\rho)\ ,
\end{split}
\end{equation}
where
\begin{equation}\label{4.5}
\nabla_\mu\phi =\dpar_\mu\phi + \CA_\mu\phi\ .
\end{equation}
The Lagrangian density \eqref{4.4} depends not only on $\tilde\CA\in\tilde\Abb$ but also on the conformal frame $\Phi\in\wt\CG$, and this should be discussed in more detail.

\noindent
{\bf Automorphisms and dressing.} In \eqref{4.4} we used the gauge potential $\tilde\CA^\Phi$ obtained by the action of the group  $\wt\CG$ on $\tilde\CA\in\tilde\Abb$,
\begin{equation}\label{4.6}
\tilde\CA\mapsto\tilde\CA^\Ups = \Ups^{-1}\tilde \CA\Ups + \Ups^{-1}\dd\Ups\quad\mbox{for}\quad\Ups\in\wt\CG\ ,
\end{equation}
by choosing $\Ups =\Phi =\rho^{-1}\phi$. Define also the right action of $\wt\CG$ on itself via
\begin{equation}\label{4.7}
\Phi\mapsto\Phi^{\hat\Ups} = \hat\Ups^{-1}\Phi\quad\mbox{for}\quad\Phi, \hat\Ups\in\wt\CG\ .
\end{equation}
Note that \eqref{4.6} and \eqref{4.7} define two independent actions of $\wt\CG$ - one action on $\tilde\Abb$ and another action on itself since in general $\Ups\ne \hat\Ups$. However, we can also consider the diagonal action $\wt\CG_{\sf diag}: \tilde\Abb\times\wt\CG\to\tilde\Abb\times\wt\CG$ by choosing $\hat\Ups =\Ups$. It is easy to see that under the action of $\wt\CG_{\sf diag}$ on pairs $(\tilde\CA , \Phi )\in\tilde\Abb^\Phi$ we have
\begin{equation}\label{4.8}
\tilde\CA^\Phi\mapsto(\tilde\CA^\Ups)^{\Phi^\Ups}=\tilde\CA^\Phi\quad\mbox{and}\quad
\wt\CF^\Phi\mapsto(\wt\CF^\Ups)^{\Phi^\Ups}=\wt\CF^\Phi\ ,
\end{equation}
i.e. $\tilde\CA^\Phi$ and $\wt\CF^\Phi$ are invariant under this action.

The group  $\wt\CG_{\sf diag}$ has subgroups  $\CG^{\sf diag}_G$ and  $\CG^{\sf diag}_{\R^+}$ .
The metric on the bundle $\wt E$ is invariant under $\CG^{\sf diag}_G$ and it is not invariant under  $\CG^{\sf diag}_{\R^+}$. Hence, the mass term \eqref{4.4} is invariant under the action of $\CG^{\sf diag}_G$ and we can define the map
\begin{equation}\label{4.9}
\tilde\Abb^\Phi \ \overset{\CG^{\sf diag}_G}{\longrightarrow}\  \tilde\Abb^\Phi /\CG^{\sf diag}_G\cong\tilde\Abb^\rho\cong\tilde\Abb\times\CG^{\sf diag}_{\R^+}\ ,
\end{equation}
obtaining that the mass term \eqref{4.4} is defined on the tangent bundle to the space $\tilde\Abb^\rho$ in \eqref{4.9}. We will omit the index word ``diag" in what follows.

\noindent
{\bf Mass term.} The $\CG_G$-invariant mass term \eqref{4.4} can be rewritten in the form
\begin{equation}\label{4.10}
\begin{split}
\CL_m^\rho &=-\sfrac12\,v^2_*\rho^2\,\tr(\nabla_\mu\phi)^\+\nabla^\mu\phi -
\sfrac{N}{2}\,v^2_*\rho^2\,(a_\mu-\dpar_\mu\ln\rho ) (a^\mu-\dpar^\mu\ln\rho )
\\
&=-\sfrac12\,v^2_*\rho^2\,\eta^{\mu\nu} \left (\tr(\CA^\phi_\mu)^\+\CA_\nu^\phi + N\,a_\mu^\rho a_\nu^\rho\right )
\ ,
\end{split}
\end{equation}
where
\begin{equation}\label{4.11}
\CA^\phi =\phi^{-1}\CA\phi + \phi^{-1}\dd\phi\quad\mbox{and}\quad a^\rho = a-\dd\ln\rho\ . 
\end{equation}
In \eqref{4.10} we can always transform $\phi$ to $\unit_N$ (unitary gauge) but this is not so for $\rho$ governed by the group $\CG^{}_{\R^+}$ acting on $\rho$, $a$ and the bundle metric. The group $\CG^{}_{\R^+}$ is not a symmetry of Lagrangians, either as a gauge group $\CG^{}_{G}$ (non-dynamic) or as Poincare group (dynamic). The map $\CA\mapsto\CA^\phi$ is called the dressing transformation, the mass term \eqref{4.3} for $\CA$ is the same as \eqref{4.10} for $\CA^\phi$, i.e. the group $\CG^{}_G$ is an artificial gauge symmetry \cite{Berg}. For the Abelian case $G=\sU(1)$ this trading of degrees of freedom between $\CA$ and scalar field $\phi\in U(1)$ was proposed by Stueckelberg~\cite{Stueck}. An overview of the dressing field method in gauge theories and many references can be found in \cite{Berg}. 

Note that for $G=U(1)$ we have $\Phi^{-1}=\rho\phi^\+=\rho e^{-\im\vph}\in\C^*=\C{\setminus}\{0\}$ and for $G=\sSU(2)$  the field $\Phi^{-1}=\rho\phi^\+\in C(\sSU(2))=C(S^3)=\Hbb{\setminus}\{0\}=\R^4{\setminus}\{0\}=\C^2{\setminus} \{0\}$ parametrizes $\C^2$-valued Higgs field on $M$. In fact, the Higgs boson can be identified with the scalar field $\rho$ whose kinetic term is given by \eqref{4.10} after choosing $a_\mu =0$. The Stueckelberg and Higgs mechanisms of gauge boson mass generation and their relationship were studied in detail in~\cite{Popov}. Here, in this article, we show that Lagrangians of scalar fields arise naturally in gauge theories with the conformal extension $\wt G$ of the structure group $G$.

\noindent
{\bf Potential for $\rho$.} We see that scalar fields in gauge theories are related with conformal frames on fibres of gauge bundles. The standard kinetic term for them is derived from the geometry of the vector bundle $\wt E$ and not entered by hand. The explicit form of the potential energy $V(\rho )$ for $\rho$ can also be derived from geometry by considering the bundle $P(M, G)=\wt P(M, \wt G)_{|\rho=1}$ of $\sSU(N)$-frames on $E=\wt E_{|\rho=1}$.

Let us fix $\rho =1$ on fibres $\wt G_x$ of $\wt P$ making it nondynamical and reducing $\wt G_x$ to $G_x$. On $G_x$ there are left-invariant one-forms $\theta^a_L$ for any $x\in M$. Let us introduce on $G_x$ a gauge potential taking value in $\frg =\rsu (N)$,
\begin{equation}\label{4.12}
A=\chi \theta^a_L I_a\ ,
\end{equation}
where $\chi$ is a real-valued function on $M=\R^{3,1}$. The curvature of $A$ is the two-form 
\begin{equation}\label{4.13}
F=\dd A + A\wedge A= (\dpar_\mu\chi )\dd x^\mu\wedge\theta^a_L I_a
+\sfrac12\chi (\chi -1)f^a_{bc} \theta^b_L\wedge\theta^c_L  I_a \ ,
\end{equation}
with components 
\begin{equation}\label{4.14}
F_{\mu a}=(\dpar_\mu\chi )I_a\quad\mbox{and}\quad F_{ab}=\chi (\chi -1)f_{ab}^cI_c\ .
\end{equation}
We choose $\chi =\rho/\rho_0$, where $\rho_0$ is some fixed function on $M$ and $\chi$ is a multiplicative ``deviation" of $\rho=\chi\rho_0$ from $\rho_0$. Then the function 
\begin{equation}\label{4.15}
\tr F^\+_{ab}F^{ab}=N\chi^2(\chi - 1)^2=\frac{N}{\rho_0^4}\,\rho^2(\rho -\rho_0)^2
\end{equation}
has extrema at
\begin{equation}\label{4.16}
\rho =0\ \ (\mbox{min})\ ,\quad \rho =\sfrac12\rho_0\ \ (\mbox{max})  \quad\mbox{and}\quad    \rho =\rho_0\ \ (\mbox{min})\ ,
\end{equation}
which are the same as extrema of the function
\begin{equation}\label{4.17}
V(\rho )=\lambda\rho^2(\rho - \rho_0)^2
\end{equation}
with $\lambda >0$. The geometric meaning of this function is that the curvature $F_{ab}$ at the minima $\rho =0$ and $\rho =\rho_0$ is equal to zero. Interestingly, this shape  \eqref{4.17} of the scalar field potential was derived in quantum gauge theory under certain assumptions \cite{Mund}. The logic used in  \eqref{4.12} -\eqref{4.17} cannot be considered as a rigorous proof, but the coincidence of  \eqref{4.17} with what is justified in \cite{Mund} makes \eqref{4.17} very plausible. Similarly, for $F_{\mu a}$ in \eqref{4.14} we have 
\begin{equation}\label{4.18}
\eta^{\mu\nu}\,\tr F^\+_{\mu a}F_{\nu a} = N\,\eta^{\mu\nu}\,\dpar_\mu\chi\dpar_\nu\chi
\end{equation}
which is proportional to the kinetic term for $\rho$ in \eqref{4.4} and \eqref{4.10} if we choose $a_\mu =\dpar_\mu\ln\rho_0$.

\noindent
{\bf Geometric Lagrangian.} We have discussed the fact that geometry of the vector bundle $\wt E$ is characterized not only by a connection $\tilde \CA\in \tilde\frg$ but also the frame field $\Phi\in\wt G$. The Lagrangian for $\tilde \CA$ has the form \eqref{4.1} and the Lagrangian for $\Phi$ is the sum of \eqref{4.10} and \eqref{4.17}. Thus, for the fields $\tilde \CA$ and $\Phi$ defining the geometry of the bundle $\wt E$, we can introduce dynamics using the Lagrangian density
\begin{equation}\label{4.19}
\begin{split}
\CL_B &=\frac{\rho^2}{4g^2_*}\,\tr (\CF_{\mu\nu}\CF^{\mu\nu})-\frac{1}{2}\,v^2_*\rho^2\,\tr(\nabla_\mu\phi)^\+\nabla^\mu\phi 
\\
&- \frac{\rho^2N}{4g^2_*}\,f_{\mu\nu}f^{\mu\nu}-
\frac{N}{2}\,v^2_*\,(\dpar_\mu\rho - a_\mu\rho ) (\dpar^\mu\rho - a^\mu\rho )-
\lambda\rho^2(\rho - \rho_0)^2\ .
\end{split}
\end{equation}
The lowest energy states (ground states) for model \eqref{4.19} are achieved on the configurations
\begin{equation}\label{4.20}
\CA^\phi =0\ \Rightarrow\ \CA = \phi^{-1}\dd\phi , \ \rho=\rho_0\quad\mbox{and}\quad a^\rho =0\  \Rightarrow\ a=\dd\ln\rho_0\ . 
\end{equation}
In general, we consider $\rho_0$ as a function of $x\in M$ and do not interpret the states \eqref{4.20} as a vacuum. The true vacuum is $\rho =0$. The function $\rho_0$ defining the ground state for $\CL_B$ depends on the gauge group under consideration, and for the group SU(3) it is different from that for the electroweak group SU(2)$\times$U(1). We will see later that $\rho_0$ is related to condensate $\langle\bar\psi\psi\rangle$ of fermion-antifermion pairs. 

Recall that the Lagrangian \eqref{4.19} is invariant under the action of the group $\CG_G=C^\infty (M,G)$, and we can always fix the unitary gauge $\phi =\unit_N$ making the field $\phi$ non-dynamical. On the other hand, the field $\rho$ is dynamical and for $G=\sSU(2)$ it is the Higgs boson, $\rho^{}_{\sSU(2)}$. In the proposed model, such a field also exists for $G=\sSU(3)$, and it is this field $\rho^{}_{\sSU(3)}$, the order parameter, that is responsible for the confinement of quarks and gluons, as will be discussed in the next section. 

In this paper, we will consider only the non-dynamic flat connection $a$, choosing $a=\dd\ln\rho_0$. Substituting $a_\mu = \dpar_\mu\ln\rho_0$ into \eqref{4.19}, we obtain
\begin{equation}\label{4.21}
\begin{split}
\CL_B (a=\dd\ln\rho_0)&=\frac{\rho_0^2}{4g^2_*}\,\chi^2\,\tr (\CF_{\mu\nu}\CF^{\mu\nu})-\sfrac{1}{2}\,v^2_*\rho_0^2\chi^2\,\tr(\nabla_\mu\phi)^\+\nabla^\mu\phi \\
&- \sfrac{N}{2}\,v^2_*\rho_0^2\dpar_{\mu}\chi\dpar^{\mu}\chi -
\lambda\rho^4_0\chi^2(\chi - 1)^2\ ,
\end{split}
\end{equation}
where $\chi :=\rho/\rho_0\ \Rightarrow\ \rho =\chi\rho_0$. Note that one can rescale $\CA\mapsto g_*\CA$ and cancel $g_*^{-2}$ term in front of the first term in \eqref{4.21} obtaining $g_*$ in the covariant derivatives as usual. 

\noindent
{\bf Running couplings.} All gauge theories are formulated in terms of bundles $E\to M$ given over the Minkowski space $M$ and having a fixed gauge coupling constant $g_*:=g_{\sf YM}$. However, one- and two-loop calculations in quantum gauge theories show that gauge couplings $g_*$ are ``running", i.e. depends on the energy scale $\mu$ (equivalently - from the distance to the source) due to vacuum polarization. The effective coupling $g_{\sf eff}$ is described by the renormalization group theory via the beta function $\beta (g_*)=\dpar g_*/\dpar\ln\mu$. In non-Abelian theories beta function is negative and $g_{\sf eff}$ decreases at high energies (small distance) and increases with decreasing energy (increasing distance). In fact,  $g_{\sf eff}$ diverges at some scale $\Lambda_c$ but this is not considered proven because $g_{\sf eff}$ is calculated perturbatively, but applying perturbation theory is no longer valid for $g_{\sf eff}>1$. At the same time, in the soliton bag models considered here, the effective coupling is running as in \eqref{4.21},
\begin{equation}\label{4.22}
\frac{1}{g^2_{\sf eff}}= \frac{1}{g^2_*}\, \rho^2\ ,
\end{equation}
already on classical or semiclassical (if quarks are quantum) level. The function $\rho$ in \eqref{4.22} tends to zero outside some region $S\subset\R^3$ (the interior of hadrons) due to equations of motion and hence $g_{\sf eff}\to\infty$ outside $S$. This is nothing but infrared slavery not based on perturbation theory.

\noindent
{\bf Fermions.} In addition to the bosons described by the Lagrangian \eqref{4.19} or \eqref{4.21}, there should be fermions in the theory. Let $\psi_b$ (``$b$" is bare) be a spinor field with values in the complex vector bundle $\wt E\to M$. It is a section of the bundle $\wt E$ tensored with the spinor bundle $W$ over $M=\R^{3,1}$. Similarly to dressed connections $\tilde\CA^\Phi$ in \eqref{4.10}, we introduce dressed fermions
\begin{equation}\label{4.23}
\psi =\Phi^{-1}\psi_b\ .
\end{equation}
The Lagrangian density for such fermionic field $\psi$ can be chosen in the form
\begin{equation}\label{4.24}
\CL_F=\rho^2\bar\psi\im\gamma^\mu(\dpar_\mu + \CA^\phi_\mu + a_\mu^\rho )\psi - (m_0 + m_1\rho )\rho^2\bar\psi\psi\ ,
\end{equation}
where $\gamma$-matrices satisfy the commutation relations $\{\gamma^\mu , \gamma^\nu\}=-2\eta^{\mu\nu}\unit_4$, $\bar\psi = \psi^\+\gamma^0$, $m_0$ and $m_1$ are real parameters (ordinary and Yukawa-type coupling). For quarks the summation over hidden flavour index is assumed. In chiral case we will take $m_0=0$ and keep the Yukawa coupling $m_1\rho$. Note that the term $(m_0 + m_1\rho )\bar\psi\psi$ was used in soliton bag models. For the full Lagrangian $\CL$ of bosons and fermions we have
\begin{equation}\label{4.25}
\begin{split}
\CL&=\CL_B +\CL_F=\frac{\rho^2}{4g^2_*}\tr (\CF_{\mu\nu}\CF^{\mu\nu})-\frac{v^2_*}{2}\rho^2\tr(\nabla_\mu\phi)^\+\nabla^\mu\phi-\frac{Nv^2_*}{2}(\dpar_\mu\rho - a_\mu\rho ) (\dpar^\mu\rho - a^\mu\rho )
\\
& - \frac{\rho^2N}{4g^2_*}\,f_{\mu\nu}f^{\mu\nu}-\lambda\rho^2(\rho - \rho_0)^2+\rho^2\bar\psi\im\gamma^\mu(\dpar_\mu + \CA^\phi_\mu + a_\mu^\rho )\psi - (m_0 + m_1\rho )\rho^2\bar\psi\psi\ ,
\end{split}
\end{equation}
where $\CA_\mu^\phi$ and $a_\mu^\rho$ are given in \eqref{4.11}.

\section{Collapsing bundles and confinement}

\noindent
{\bf Research proposals}. Although the word ``confinement" appears in the title of this paper, this topic is not the main one. What we really want to understand is how scalar fields enter the theory as part of the geometry of gauge bundle rather then being introduced by hand. It is assumed that scalar fields not only give masses for gauge bosons, but under certain conditions are also responsible for confinement.

In the conventional approach, QCD is described in terms of Hermitian vector $\C^3$-bundle $E$ over Minkowski space $M$. Gluons are connections $\CA\in\rsu(3)$ on this bundle $E$ of color degrees of freedom and quarks $\psi$ are $\C^3$-valued spinors,  sections of $E$ tensored with the spinor bundle $W$ over $M=\R^{3,1}$. It is considered that the bundle $E$ is given over the {\it whole} space $\R^3$ and for any moment of time. Therefore, gluons $\CA$ and quarks $\psi$ are also defined over the whole Minkowski space $M$. It is also assumed that the bundle $E$ is given even if there are no quarks and gluons, i.e. when $\CA =0=\psi$. Then, in the standard approach, it is assumed that the fields of quarks $\psi$ and gluons $\CA$ are non-zero inside the region $S$ (the interior of hadrons) and disappear outside $S\subset\R^3$ either due to some properties of QCD vacuum that have not yet been proven or due to some properties of QCD in the infrared region that also have not yet been proven. However, none of the proposed confinement scenario (dual superconductivity, Green's function approach, Gribov-Zwanziger scenario, etc.) has been convincingly substantiated despite 50 years of efforts.

We repeat that everything said in the previous paragraph refer to the standard Lagrangian of quantum chromodynamics given for the bundle $E$, which exists on the entire Minkowski space $M$ even if there are no quarks and gluons. However, bag models say that bundle $E$ appears only if quarks $\psi$ are inserted in some region $S$ of space $\R^3$ and all fields outside $S$ disappear due to the fact that the field $\rho$ entering the Lagrangian \eqref{4.25} tends to zero outside $S$. In the previous sections, we showed that the scalar field $\rho$ is the conformal factor of metrics on the fibres $\wt E_x$ of the bundle $\wt E$ and, therefore, when $\rho\to 0$, the metric on fibres is scaled down and the bundle $\wt E$ collapses into Minkowski space outside $S\subset\R^3$. It is obvious that $\CA =0=\psi$ if $\rho =0$. The state $\rho =0$ is the true vacuum. In other words, when quantizing, one should introduce the creation and annihilation operators  not only for quarks and gluons, but also for bundles $\wt E$ (using $\rho$ and $\dpar_t\rho$). In what follows, we will focus on how the creation and annihilation of bundles can be described in terms of geometry.

\noindent
{\bf Collapsing bundles}. To illustrate what was said above about the collapse of bundles, consider for example the Hopf bundle
\begin{equation}\label{5.1}
S^3 \ \overset{S^1}{\longrightarrow}\  S^2\ ,
\end{equation}
which is the principal bundle $S^3=P(S^2, \sU(1))$ over $S^2$ with $\sU(1)$ as fibres, and the associated complex vector bundle
\begin{equation}\label{5.2}
E_\C \ \overset{\C}{\longrightarrow}\  S^2\ ,
\end{equation}
for $E_\C = P\times_{\sU(1)}\C$.  These two bundles model our bundles $\wt P$ and $\wt E$.

According to Cheeger and Gromov \cite{Gromov}, the collapse of bundles was first considered by M.Berger in 1962 on the example of the collapse of $S^3$, obtained by shrinking the circle $S^1$ of the Hopf fibration  \eqref{5.1} and the limit of this collapse is the sphere $S^2$. The metric on the sphere $S^2\cong\C P^1$ has the form
\begin{equation}\label{5.3}
\dd s^2_{\CPP^1}=\frac{4R^2\dd\zeta\dd\bar\zeta}{(1+\zeta\bar\zeta)^2}\ ,
\end{equation}
where $R$ is the radius of $S^2$ and $\zeta\in\C$ is a local complex coordinate. On the Hopf bundle \eqref{5.1} there is the unique $\sSU(2)$-invariant connection $a$, having in the dimensionless coordinates $\zeta , \bar\zeta$ the form
\begin{equation}\label{5.4}
a=\frac{1}{2(1+\zeta\bar\zeta)} (\bar\zeta\dd\zeta - \zeta\dd\bar\zeta )\in {\rm u}(1)\ .
\end{equation}
The curvature of this connection is
\begin{equation}\label{5.5}
f=\dd a =-\frac{\dd\zeta\wedge\dd\bar\zeta}{(1+\zeta\bar\zeta)^2}
\end{equation}
and the $\sU(1)$ Yang-Mills equations on $S^2$ are satisfied. On the associated complex vector bundle \eqref{5.2} we have the same connection $a$ and curvature $f$.

Let $\vph$ be a coordinate on fibres $\sU(1)$ (circles $S^1$ of radius $R$) of the bundle \eqref{5.1}  and $z$ be a complex coordinate on fibres $\C$ of the bundle \eqref{5.2}. As shown in Section 3, the metrics on the total spaces of bundles \eqref{5.1} and \eqref{5.2} are
\begin{equation}\label{5.6}
\dd s^2_P = \dd s^2_{S^2} + R^2(\dd\vph -\im a)^2\ ,
\end{equation}
\begin{equation}\label{5.7}
\dd s^2_{E_\C} = \dd s^2_{S^2} + (\dd z + az)(\dd\bar z - a\bar z)\ ,
\end{equation}
where $\dd s^2_{S^2}$ and $a$ are given in \eqref{5.3} and \eqref{5.4}. Note that for $z=R\exp(\im\vph)$ metric \eqref{5.7} is reduced to \eqref{5.6}. According to M.Berger, we can deform these metrics by multiplying the metrics on fibres by a parameter $\veps^2$,
\begin{equation}\label{5.8}
\dd s^2_{P^\veps} = \dd s^2_{S^2} + \veps^2R^2(\dd\vph -\im a)^2= \dd s^2_{S^2} + R^2(\veps\dd\vph -\im\veps a)^2\ ,
\end{equation}
\begin{equation}\label{5.9}
\dd s^2_{E_\C^\veps} =\dd s^2_{S^2} + \veps^2\Theta^z\bar\Theta^{\bar z}=\dd s^2_{S^2} + (\veps\dd z + \veps az)(\veps\dd\bar z -\veps a\bar z)\ .
\end{equation}
These deformations mean that the radius of the circle $S^1_\veps$ in fibres of \eqref{5.8} is $R_\veps =\veps R$ and for $\veps\to 0$ the deformed metrics \eqref{5.8} and \eqref{5.9} are reduced to the metric \eqref{5.3} on $S^2$. In addition, from these formulae we see that the deformed connection on the bundles \eqref{5.1} and \eqref{5.2} is $a^\veps =\veps a$ and $a^\veps \to 0$ for $\veps\to 0$ regardless of any equations. Similarly, sections $\psi^\veps$ of the bundle $E_\C^\veps$ tends to zero for $\veps\to 0$. And this is a common characteristic feature of collapsing bundles. If we now replace $S^2$ by Minkowski space $M$ and the conformal factor $\veps^2$ by a function $\rho^2$, we obtain the gauge theory considered in the previous section. If the function $\rho$ will vanish outside some region $S$ in $\R^3$, then the bundle $\wt P(M, \wt G)$ and $\wt E(M,V)$ will collapse to Minkowski space outside $S$.

\noindent
{\bf $\sSU(3)$-bundles and confinement}. To describe gluons and quarks, we use Lagrangian \eqref{4.25} in the form
\begin{equation}\label{5.10}
\begin{split}
\CL_{\sSU(3)}&=\frac{1}{4g^2_s}\,\rho^2\tr (\CF_{\mu\nu}\CF^{\mu\nu})+
\sfrac{1}{2}\,v^2_s\rho^2\,\eta^{\mu\nu}\tr\CA_\mu^\phi\CA_\nu^\phi 
- \frac{3}{4g^2_s}\,\rho^2f_{\mu\nu}f^{\mu\nu}-
\sfrac{3}{2}\,v^2_s\,\rho^2\,\eta^{\mu\nu}\, a_\mu^\rho\, a_\nu^\rho 
\\
&-\lambda\rho^2(\rho - \rho_0)^2+\rho^2\bar\psi\im\gamma^\mu(\dpar_\mu + \CA^\phi_\mu + a_\mu^\rho )\psi - (m_0 + m_1\rho )\rho^2\bar\psi\psi\ ,
\end{split}
\end{equation}
where $\CA^\phi$ and $a^\rho$ are the dressed fields given in \eqref{4.11} and $g_s$ is a gauge coupling for $G=\sSU(3)$. The last term in \eqref{5.10} contains the mass term $m_0$ and the Yukawa coupling $m_1$ as in the Friedberg-Lee model \cite{FrLee}. Chiral symmetry is broken, we consider infrared region.

We want to understand the origin of the potential energy for the scalar field $\rho$,
\begin{equation}\label{5.11}
V(\rho ) = \lambda\rho^2(\rho - \rho_0)^2 = \lambda(\rho^4 - 2\rho_0\rho^3 + \rho_0^2\rho^2)\ ,
\end{equation}
or, more precisely, to find out the origin of $\rho_0$. To do this, we consider \eqref{5.10} without gauge fields, assuming
\begin{equation}\label{5.12}
\CA^\phi =0\ ,\quad a^\rho =0 \quad\mbox{and}\quad  \rho_0=0
\end{equation}
and obtaining
\begin{equation}\label{5.13}
-\frac{1}{\lambda}\,\CL_{\sSU(3)}=\rho^4 -2\left(-\frac{m_1}{2\lambda}\bar\psi\psi\right)\rho^3+
\left(-\frac{1}{\lambda}\bar\psi\,\im\,\gamma^\mu\dpar_\mu\psi + \frac{m_0}{\lambda}\bar\psi\,\psi\right)\rho^2\ .
\end{equation}
In the absence of gluons and quarks, i.e. \eqref{5.13} with $\psi =0$, the Lagrangian extremum and the potential energy minimum are reached at
\begin{equation}\label{5.14}
\rho =0\ ,
\end{equation}
which is the true vacuum.

We now assume that the quark field $\psi$ is nonzero and compare formulae  \eqref{5.11} and \eqref{5.13}. Let us introduce a composite field
\begin{equation}\label{5.15}
\rho_0 =-\frac{m_1}{2\lambda}\bar\psi\psi
\end{equation}
and assume that $\psi$ is a solution to the nonlinear Dirac equation
\begin{equation}\label{5.16}
\im\,\gamma^\mu\dpar_\mu\psi - m_0\psi+ \frac{m_1^2}{4\lambda}(\bar\psi\,\psi)\psi=0\ .
\end{equation}
Then, on solutions of eq.\eqref{5.16}, the Lagrangian density \eqref{5.13} becomes
\begin{equation}\label{5.17}
\CL_{\sSU(3)} (\CA^\phi{=}a^\rho{=}0, \psi) = -\lambda\rho^2(\rho - \rho_0)^2=-V(\rho )\ ,
\end{equation}
where $\rho_0$ is given by \eqref{5.15}.

The nonlinear Dirac equation \eqref{5.16} is called Soler equation \cite{Soler}. This equation has stable solutions of the form
\begin{equation}\label{5.18}
\psi_S=e^{-\im\omega t}\vph_S(x)\quad\mbox{for}\quad \omega\in\R^+\quad\mbox{and}\quad x\in\R^3\ ,
\end{equation}
where $\vph_S(x)$ is localized around $x=0$ and have finite energy only if
\begin{equation}\label{5.19}
0<\omega <m_0\ .
\end{equation}
The above $\psi_S$ in \eqref{5.16}-\eqref{5.18} is a soliton type field with finite energy supported on $S\in \R^3$. The Soler equation \eqref{5.16} and its particle-like solutions of the form \eqref{5.18} have been intensively studied in the mathematical literature (see e.g. \cite{CV, Est, Cue} and references therein). We see that solutions of this nonlinear Dirac equation are directly related to the description of hadrons.

Having found $\rho_0$ from \eqref{5.15}-\eqref{5.19}, we next consider excitations of the form 
\begin{equation}\label{5.20}
\rho = \chi\rho_0\quad\mbox{and}\quad a=\dd\ln\rho_0\quad\Rightarrow\quad a^\rho =a-\dd\ln\rho=-\dd\ln\chi \ .
\end{equation}
Substituting \eqref{5.20} into \eqref{5.10}, we obtain the Lagrangian density
\begin{equation}\label{5.21}
\begin{split}
\CL_{\sSU(3)}(a&=\dd\ln\rho_0)=\frac{1}{4g^2_s}\,\rho_0^2\chi^2\,\tr (\CF_{\mu\nu}\CF^{\mu\nu})-
\sfrac{1}{2}\,\rho_0^2v^2_s\,\chi^2\,\tr(\nabla_\mu\phi)^\+\nabla^\mu\phi 
- \sfrac{3}{2}\rho_0^2v^2_s\,\dpar_{\mu}\chi \dpar^{\mu}\chi \\
&-
\lambda\rho_0^4\chi^2(\chi - 1)^2 +\rho_0^2\chi^2\bar\psi\im\gamma^\mu(\dpar_\mu + \CA^\phi_\mu -\dpar_\mu\ln\chi )\psi - (m_0 + m_1\rho_0 \chi)\rho_0^2\chi^2\bar\psi\psi\ ,
\end{split}
\end{equation}
where $\psi$ is considered as an excitation over a solution $\psi_S$ from \eqref{5.16}-\eqref{5.19}. Localized solutions of equations \eqref{5.16} can explain confinement in theory \eqref{5.21}.

\noindent
{\bf $\sSU(2)_L$-bundles and masses.} We now consider a Lagrangian of type \eqref{4.25} for the group $\sSU(2)_L\times\sU(1)$ of electroweak interaction. Note that the group $\sU(1)$ in $\sSU(2)_L\times\sU(1)$ will not play a role in what we will discuss below. Therefore, to simplify the discussion, we put the charge and the corresponding Abelian gauge field equal to zero, leaving only the chiral group $G=\sSU(2)_L$  and its conformal extension $\wt G$ with the gauge coupling $g_*=g_w$. In this subsection we will consider the lepton sector. 

Consider the principal bundle $\wt P(M, \wt\sSU(2))$ and the associated complex vector bundle $\wt E_{\C^2}$ with $\C^2$-fibres discussed in Section 3. Let $W\to M$ be a $\C^4$-vector bundle of Dirac spinors on Minkowski space $M$. Leptons are groupped into three families (generations), so that $\psi =(\nu\ l)^\top$ are sections of the bundle $\wt E_{\C^2}\times W$, where $l=e$, $\mu$ or $\tau$ (charged leptons) and $\nu =\nu_e$, $\nu_\mu$ or $\nu_\tau$ (neitrinos). We choose
\begin{equation}\label{5.22}
\gamma^5=\begin{pmatrix}-\unit_2 & 0\\0 & \unit_2\end{pmatrix},\ \psi^{}_L=\sfrac12(1-\gamma^5)\psi ,\ \psi^{}_R=\sfrac12(1+\gamma^5)\psi ,\ 
\psi^{}_L=\begin{pmatrix}\nu^{}_L\\ l^{}_L\end{pmatrix}\ \mbox{and}\ \psi^{}_R=\begin{pmatrix}\nu^{}_R\\ l^{}_R\end{pmatrix}.
\end{equation}
We keep in $\psi_R$ the right-handed neitrino $\nu_R$ since this does not affect further consideration. If desired, one can set $\nu_R=0$.

The splitting \eqref{5.22} into left-handed and right-handed spinors corresponds to the splitting of the bundle $\wt E_{\C^2}\times W=\wt E_{\C^2}{\otimes}W_L\oplus\wt E_{\C^2}{\otimes}W_R$ so that $\psi_L\in\wt E_{\C^2}\times W_L$ and $\psi_R\in\wt E_{\C^2}\times W_R$. Accordingly, we will identify $\sSU(2)_L$ as the group
\begin{equation}\nonumber
G=\sSU(2)_L=\begin{pmatrix}\sSU(2)&0\\0&\unit_2\end{pmatrix}\ni g_L=\begin{pmatrix}g&0\\0&\unit_2\end{pmatrix}
\end{equation}
so that $G$ acts only on $\psi_L$ (doublets) and $\psi^{}_R$ is a singlet. However, the conformal extension of $G$ is defined as $\tilde g_L^\+\tilde g_L =\rho^2\unit_4$ and therefore the scaling group $\CG^{}_{\R^+}$ is not chiral, it acts both on $\psi_L$ and $\psi^{}_R$. Thus, the covariant derivative of $\psi =\psi_L\oplus\psi_R$ has the form
\begin{equation}\label{5.23}
\nabla^L_\mu\psi = \begin{pmatrix}(\dpar_\mu + \tilde\CA_\mu^{\tilde\phi}+ \tilde a_\mu^{\tilde\rho})\psi^{}_L\\    (\dpar_\mu + \tilde a_\mu^{\tilde\rho})\psi^{}_R \end{pmatrix}
\quad \mbox{for}\quad\psi=\begin{pmatrix}\psi^{}_L\\ \psi^{}_R\end{pmatrix}\ ,
\end{equation}
where we denoted by tilde the gauge field $\tilde\CA\in\rsu(2)$, scale connection $\tilde a\in\R$ and scalars $\tilde\phi\in\sSU(2)$, $\tilde\rho\in\R^+$. Note that $\tilde\rho$ (Higgs boson) from $\sSU(2)_L$-theory is not related to $\rho$ from the considered $\sSU(3)$-theory, these are different scalar fields.

The conventional fermion mass term $\wt m_0\bar\psi\psi = \wt m_0(\psi^{\+}_L\psi^{}_R + \psi^{\+}_R\psi^{}_L)$ is forbidden since such term does not respect $\sSU(2)_L$ gauge symmetry. Hence, the fermion masses in electroweak sector result only from Yukawa-type interaction which in the unitary gauge has the form $\wt m_1\tilde\rho\bar\psi\psi$. The Lagrangian density in the unitary gauge $\tilde\phi=1$ has the form
\begin{equation}\label{5.24}
\begin{split}
\CL_{\sSU(2)_L}&=\frac{1}{4g^2_w}\,\tilde\rho^2\tr (\wt\CF_{\mu\nu}\wt\CF^{\mu\nu})+
\frac{1}{2} v^2_w\,\tilde\rho^2\tr(\tilde\CA_\mu\tilde\CA^\mu)- \frac{1}{2g^2_w}\,\tilde\rho^2\tilde f_{\mu\nu}\tilde f^{\mu\nu}
\\
& - v^2_w\tilde\rho^2\eta^{\mu\nu}\, \tilde a_\mu^{\tilde\rho}\, \tilde a_\nu^{\tilde\rho}- \tilde\lambda \tilde\rho^2(\tilde\rho - \tilde\rho_0)^2
+\tilde\rho^2\bar\psi\im\gamma^\mu\nabla^L_\mu \psi - \tilde m_1\tilde\rho^3 \bar\psi\psi \ ,
\end{split}
\end{equation}
where $\wt\CF_{\mu\nu}\in\rsu(2)$ is the curvature of $\tilde\CA_\mu\in\rsu(2)$ and $g_w$ is a gauge coupling for $G=\sSU(2)_L$.

Now we apply the same logic as in the $\sSU(3)$ case. We assume that in the absence of gauge and fermionic fields, the potential has the form $\tilde\lambda\tilde\rho^4$ with $\tilde\rho_0=0$ and $\tilde\rho_0$ arises as a ``condensate" 
\begin{equation}\label{5.25}
\tilde\rho_0=-\frac{\wt m_1}{2\tilde\lambda}\bar\psi\psi\ ,
\end{equation}
where $\psi$ is a solution of the equation
\begin{equation}\label{5.26}
\im\gamma^\mu\dpar_\mu\psi +\frac{\wt m_1^2}{4\tilde\lambda}(\bar\psi\psi )\psi =0\ .
\end{equation}
This is the Soler equation \eqref{5.16} with zero mass $\wt m_0=0$. It is known that eq.\eqref{5.26} with $\wt m_0=0$ cannot have localized solutions with finite energy (see e.g. \cite{CV, Est, Cue} for a proof and discussion). However, it has plane-wave solutions
\begin{equation}\label{5.27}
\psi = e^{\im p_\mu x^\mu}\psi_0
\end{equation}
and one can choose $\psi_0$ in \eqref{5.27} such that $\tilde\rho_0=1$ in \eqref{5.25}. This seems natural, since electrons can move throughout space, i.e. their field $\psi$ is given on the whole Minkowski space $\R^{3,1}$.

Thus, the chiral nature of weak interaction with the group $\sSU(2)_L$ leads to the masses of $\sSU(2)_L$ gauge bosons and fermions, and not to confinement. Having found $\tilde\rho_0=1$ we choose
\begin{equation}\label{5.28}
\tilde a =\dd\ln\tilde\rho_0=0\quad \mbox{for}\quad \tilde\rho_0=1
\end{equation}
and $\tilde\rho =\chi\tilde\rho_0=\chi$. Substituting \eqref{5.28} into \eqref{5.24}, we obtain
\begin{equation}\label{5.29}
\begin{split}
\CL_{\sSU(2)_L}
&=\frac{1}{4g^2_w}\,\chi^2\,\tr (\wt\CF_{\mu\nu}\wt\CF^{\mu\nu})+
\sfrac12\, v^2_w\,\chi^2\,\tr(\tilde\CA_\mu\tilde\CA^\mu)- v^2_w\dpar_\mu{\chi}\, \dpar^\mu{\chi}\\
& - \tilde\lambda\chi^2(\chi - 1)^2
+\chi^2(\bar\psi\im\gamma^\mu\nabla_\mu^L \psi - \tilde m_1\chi \bar\psi\psi )\ ,
\end{split}
\end{equation}
where $\chi$ is close to unity, $\chi =1+h$, $h\ll 1.$ This is the Lagrangian in unitary gauge with broken $\sSU(2)_L$.

\noindent
{\bf $\sSU(2)_L$ and quarks.} In the electroweak case, leptons are sections of the bundle $\wt E_{\C^2}\otimes\wt E_{\C}\otimes W$ associated with the group $\sSU(2)_L\times\sU(1)$. It is known that the Higgs scale $R_w$ is much smaller than the confinement scale $R_s$ (we mean length scale) and Lagrangian \eqref{5.10} was considered for distances where $\sSU(2)_L$ is broken with $\tilde\rho = \tilde\rho_0=1$. In the $\sSU(3)$ case described by \eqref{5.10}, we considered quarks $\psi$ as sections of the bundle $\wt E_{\C^3}^{}$ associated with the color group $\sSU(3)$. This is correct. However, in the general case, quark doublets are sections of the bundle $\hat E:=\wt E_{\C^3}\otimes\wt E_{\C^2}\otimes\wt E_{\C}\otimes W$, and at length scale of order $R_w$ or less, the conformal factor for the metric on fibres of the bundle $\hat E$ will be
\begin{equation}\label{5.30}
\hat\rho^2=\rho^2\tilde\rho^2\ ,
\end{equation}
where $\rho^2$ is a conformal factor on $\wt E_{\C^3}$ and $\tilde\rho^2$ is a conformal factor on $\wt E_{\C^2}\otimes\wt E_{\C}$. Both factors were considered separately - for quarks in \eqref{5.10} and for leptons in \eqref{5.24}.

Lagrangian density for quark doublets at distances $r\le R_w$ can be written as
\begin{equation}\label{5.31}
\CL_{F}=\tilde\rho^2\rho^2\bar\psi\im\gamma^\mu\hat\nabla_\mu \psi - \tilde\rho^3(m_0 + m_1\rho)\rho^2\bar\psi\psi\ ,
\end{equation}
where $\hat\nabla_\mu$ is a combination of $\sSU(3)$ and $\sSU(2)_L$ covariant derivatives. Also now we need to consider the potential
\begin{equation}\label{5.32}
V(\rho , \tilde\rho )=\tilde\lambda\tilde\rho^2(\tilde\rho - \tilde\rho_0)^2 + \lambda\rho^2(\rho - \rho_0)^2
\end{equation}
for the fields $\tilde\rho , \rho$. Note that for $r\le R_w$ the function $\rho$ is practically constant (it is also large, and hence $g_{\sf eff}=\rho^{-1}g_s$ is small) and such $\rho$ can be inserted into $\psi$, so that $m_0+m_1\rho$ can be identified with $\wt m_1$ in \eqref{5.24}-\eqref{5.29}. Therefore at very short distances quarks behave like free fields (asymptotic freedom) from \eqref{5.27} and from \eqref{5.25}-\eqref{5.28} we obtain $\tilde\rho_0=1$ for these distances. Then, on the confinement scale $R_s\gg R_w$, we find ourselves in the situation described by \eqref{5.10} with $\tilde\rho=\tilde\rho_0=1$ and arrive at equations \eqref{5.13}-\eqref{5.21}. Confinement of quarks and gluons emerges {\it after} electroweak symmetry breaking $\sSU(2)_L\times\sU(1)\to\sU(1)$. Thus, the field $\tilde\rho$ is responsible for the masses, and the field $\rho$ for the confinement.

\section{Conclusions}

The equations of electrodynamics were introduced by Maxwell as a result of summing up all the accumulated experience in the study of electrical and magnetic phenomena. On the other hand, Yang and Mills introduced their equations as a formal generalization of Maxwell's equations, without connection to experiments. Already at the moment of introduction, their equations contradicted the experience indicating the absence of massless non-Abelian gauge bosons, which provoked sharp criticism from Pauli. This problem was partially solved using the Higgs mechanism, but after the introduction of quarks and QCD, the confinement problem arose, which was not solved within the framework of Yang-Mills theory for 50 years. In fact, the Yang-Mills model inherits the masslessness and lack of confinement of Maxwell's theory due to a too straightforward generalization of the Abelian case. To resolve this problem, one needs to modify and generalize the model, putting confinement into it as input data, at the level of first principles, and not as a property that suddenly miraculously turns out after some external manipulations.

In this paper, we proposed a program for including scalar fields in gauge theories as part of the geometry of vector bundles. The proposed modification of Yang-Mills theory is aimed at improving the understanding of the nature of scalar fields (are they matter or geometry?), confinement and other nonperturbative effects. The proposed geometric introduction of scalar fields through a conformal extension $\wt G$ of gauge groups $G$ makes it possible to describe not only the generation of masses through interaction with a scalar field $\tilde\rho$ in the electroweak $\sSU(2)_L\times\sU(1)$ theory, but also the confinement generated by a compactly supported scalar field $\rho$ in $\sSU(3)$ theory. These scalar fields are dynamical and their vanishing means the collapse of the vector bundle $\wt E\to M$ into Minkowski space. This means that, when quantized, the scalar field $\rho$ will give operators of creation and annihilation of bundles $\wt E$. In other words, if there is no matter $\psi$ (quarks and leptons), then there is no geometry (bundles $\wt E$ and connections $\tilde\CA$). Accordingly, the vacuum is characterized not only by the absence of gauge and fermionic fields, i.e. $\tilde\CA =0$ and $\psi=0$, but also by the absence of the scalar field $\rho$. The vanishing of $\rho$ means the collapse of the gauge bundle to Minkowski space.

\bigskip 

\noindent {\bf Acknowledgments}

\noindent
I am grateful to Tatiana Ivanova for stimulating discussions.
This work was supported by the Deutsche Forschungsgemeinschaft.

\bigskip 


\end{document}